\documentclass[3p]{elsarticle}

\usepackage{amsmath}
\usepackage{amssymb}
\usepackage{graphicx}
\usepackage{makeidx}  
\usepackage{enumerate}
\usepackage{textcomp}         
\usepackage{multirow}
\usepackage{multicol}
\usepackage{verbatim}
\usepackage[boxed]{algorithm2e}
\usepackage[caption=false]{subfig}
\usepackage{url}

\makeatletter
\def\@author#1{\g@addto@macro\elsauthors{\normalsize%
    \def\baselinestretch{1}%
    \upshape\authorsep#1\unskip\textsuperscript{%
      \ifx\@fnmark\@empty\else\unskip\sep\@fnmark\let\sep=,\fi
      \ifx\@corref\@empty\else\unskip\sep\@corref\let\sep=,\fi
      }%
    \def\authorsep{\unskip,\space}%
    \global\let\@fnmark\@empty
    \global\let\@corref\@empty  
    \global\let\sep\@empty}%
    \@eadauthor={#1}
}
\makeatother

\begin{document}
\begin{frontmatter}
\title{Symmetry Breaking Predicates for SAT-based DFA Identification}

\author{Vladimir Ulyantsev\corref{maincor}}
\ead{ulyantsev@rain.ifmo.ru}

\author{Ilya Zakirzyanov\corref{cor}}
\ead{zakirzyanov@rain.ifmo.ru}

\author{Anatoly Shalyto}
\ead{shalyto@mail.ifmo.ru}

\address{ITMO University, Kronverkskiy avenue 49, Saint-Petersburg, Russia}

\cortext[maincor]{Principal corresponding author}
\cortext[cor]{Corresponding author}

\begin{abstract}
  It was shown before that the NP-hard problem of deterministic finite automata (DFA) 
  identification can be effectively translated to Boolean satisfiability (SAT). 
  Modern SAT-solvers can tackle hard DFA identification instances efficiently. 
  We present a technique to reduce the problem search space by enforcing an enumeration of DFA states 
  in depth-first search (DFS) or breadth-first search (BFS) order. We propose symmetry breaking predicates, 
  which can be added to Boolean formulae representing various DFA identification problems. 
  We show how to apply this technique to DFA identification from both noiseless and noisy data.
  Also we propose a method to identify all automata of the desired size.
  The proposed approach outperforms the current state-of-the-art DFASAT method for DFA identification from
  noiseless data. A big advantage of the proposed approach is that it allows to determine exactly the existence or 
  non-existence of a solution of the noisy DFA identification problem unlike metaheuristic approaches such as genetic algorithms.
\end{abstract}


\end{frontmatter}

\section{Introduction}

Deterministic finite automata (DFA) are models that recognize regular languages~\cite{hmu2006},
therefore the problem of DFA identification (induction, learning) is 
one of the best studied~\cite{higuera} in grammatical inference. 
The identification problem consists of finding a DFA with the minimal number of states that is 
consistent with a given set of strings with language attribution labels.
This means that such a DFA rejects the negative example strings and accepts the positive example strings. 
It was shown in~\cite{gold1978} that finding a DFA with a given upper bound on its size (number of states) is an NP-complete problem.
Besides, in~\cite{pitt1993} it was shown that this problem cannot be approximated within any polynomial.

Despite this theoretical difficulty, several efficient DFA identification algorithms exist~\cite{higuera}. 
The most common approach is the evidence driven state-merging (EDSM) algorithm~\cite{abbadingo1998}. 
The key idea of this algorithm is to first construct an augmented prefix tree acceptor (APTA), 
a tree-shaped automaton, from the given labeled strings, 
and then to apply iteratively a state-merging procedure until no valid merges are left.
Thus EDSM is a polynomial-time greedy method that tries to find a good local optimum. 
EDSM participated in the Abbadingo DFA learning competition~\cite{abbadingo1998} and won it (in a tie). 
To improve the EDSM algorithm several specialized search procedures were proposed, see, e.g.,~\cite{lang99,bugalho2005}.
One of the most successful approaches is the EDSM algorithm in the red-blue framework~\cite{abbadingo1998}, 
also called the Blue-fringe algorithm.

The second approach for DFA learning is based on evolutionary computation; early work includes~\cite{dupont1994,luke1999}. 
Later the authors of~\cite{lucas2003} presented an effective scheme for evolving DFA with a multi-start random hill climber, 
which was used to optimize the transition matrix of the identified DFA.
A so-called smart state labeling scheme was applied to choose state labels optimally, 
given the transition matrix and the training set. 
The authors emphasized that smart selection of state labels gave the evolutionary method a 
significant boost which allowed it to compete with EDSM.
Authors find that the proposed evolutionary algorithm (EA) outperforms the EDSM algorithm on small target DFA 
when the training set is sparse. 
For larger automata with $32$ states, the hill climber fails and EDSM clearly outperforms it.

The challenge of the GECCO 2004 Noisy DFA competition~\cite{gecco2004} was to learn the target DFA when 
$10$ percent of the given training string labels had been randomly flipped. 
In~\cite{lucas2005} Lucas and Reynolds show that within limited time the EA with smart state 
labeling is able to identify the target DFA even at such high noise level. 
The authors compared their algorithm with the results of the GECCO competition and found 
that the EA clearly outperformed all the entries. 
Thereby it is the state-of-the-art technique for learning DFA from noisy training data.

In several cases the best solution for noiseless DFA identification is the \textit{translation-to-SAT} technique~\cite{heule2010}, 
which was altered to suit the \textit{StaMInA} (State Machine Inference Approaches) 
competition~\cite{walkinshaw2013} and ultimately won. 
The main idea of that algorithm is to translate the DFA identification problem to Boolean satisfiability (SAT).
Thus it is possible to use highly optimized modern DPLL-style SAT solving techniques~\cite{biere2009}.
The translation-to-SAT approach was also used to efficiently tackle problems such as bounded model checking~\cite{amla2005analysis},
solving SQL constraints by incremental translation~\cite{lohfert2008}, 
analysis of JML-annotated Java sequential programs~\cite{galeotti2013},
extended finite-state machine induction~\cite{ulyantsev2011}.

Many optimization problems exhibit symmetries~-- 
groups of solutions which can be obtained from each other via some simple transformations. 
To speed up the solution search process we can reduce the problem search space by performing \textit{symmetry breaking}.
In DFA identification problems the most straightforward symmetries are groups of isomorphic automata.
The idea of avoiding isomorphic DFAs by fixing state numbers in 
breadth-first search (BFS) order was used in the state-merging approach~\cite{lambeau2008}
(function \texttt{NatOrder}) and in the genetic algorithm from~\cite{chambers2010} (\textit{Move To Front} reorganization).
Besides, in~\cite{heule2010} symmetry breaking was performed by fixing some colors of the APTA vertices from 
a clique provided by a greedy \textit{max-clique} algorithm
applied in a preprocessing step of translation-to-SAT technique.

In this paper we propose new symmetry breaking predicates~\cite{biere2009} which can be added to 
Boolean formulae representing various DFA identification problems.
These predicates enforce DFA states to be enumerated in the DFS (depth-first search) or BFS order.
This approach clearly outperforms current state-of-the-art DFASAT from~\cite{heule2010}.
The proposed predicates cannot be applied with the max-clique technique~\cite{heule2010} at the same time, 
but our approach is more flexible.
To show the flexibility of the approach, we pay our attention to the case of noisy DFA identfication.
Therefore we propose a modification of the noiseless translation-to-SAT for the noisy case (Section~\ref{noisy}).
We show that the previously proposed max-clique technique is not applicable in this case while our BFS-based approach is.
The big advantage of our approach is that we can determine the existence or non-existence of a solution in this case unlike genetic algorithms. 
Experiments showed that using BFS-based symmetry breaking predicates can significantly reduce the time of
algorithm execution. We also show that our strategy outperforms the current state-of-the-art EA from~\cite{lucas2005} if the number 
of the target DFA states, the noise level and the number of strings are small.
We also propose a modification of this method to solve the problem of finding all automata with the minimal number of states which are 
consistent with a given set of strings.


\section{Encoding DFA Identification into SAT}
  \label{exact-DFA}

\begin{figure}
  \centering
  \includegraphics[width=2.3in]{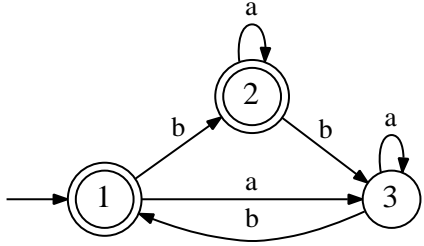}
  \caption{An example of a DFA}
  \label{dfa-example}
\end{figure}

The goal of DFA identification is to find a smallest DFA $A$ such that every string from $S_+$, a set of  
positive examples, is accepted by $A$, and every string from $S_-$, a set of negative examples, is rejected. 
The size of $A$ is defined as the number of states $C$ it has.
The alphabet $\Sigma = \{l_1, \ldots, l_L\}$ of the sought DFA $A$ is the set of all symbols from $S_+$ and $S_-$ where $L$ is the alphabet size.
The example of the smallest DFA for $S_+=\{ab, b, ba, bbb\}$ and $S_-=\{abbb, baba\}$ is shown in Fig.~\ref{dfa-example}.
In this work we assume that DFA states are numbered from $1$ to $C$ and the start state has number $1$.

\begin{figure}[!ht]
  \subfloat[An example of an APTA for $S_+=\{ab, b, ba, bbb\}$ and $S_-=\{abbb, baba\}$\label{APTA}]{%
    \includegraphics[width=3in]{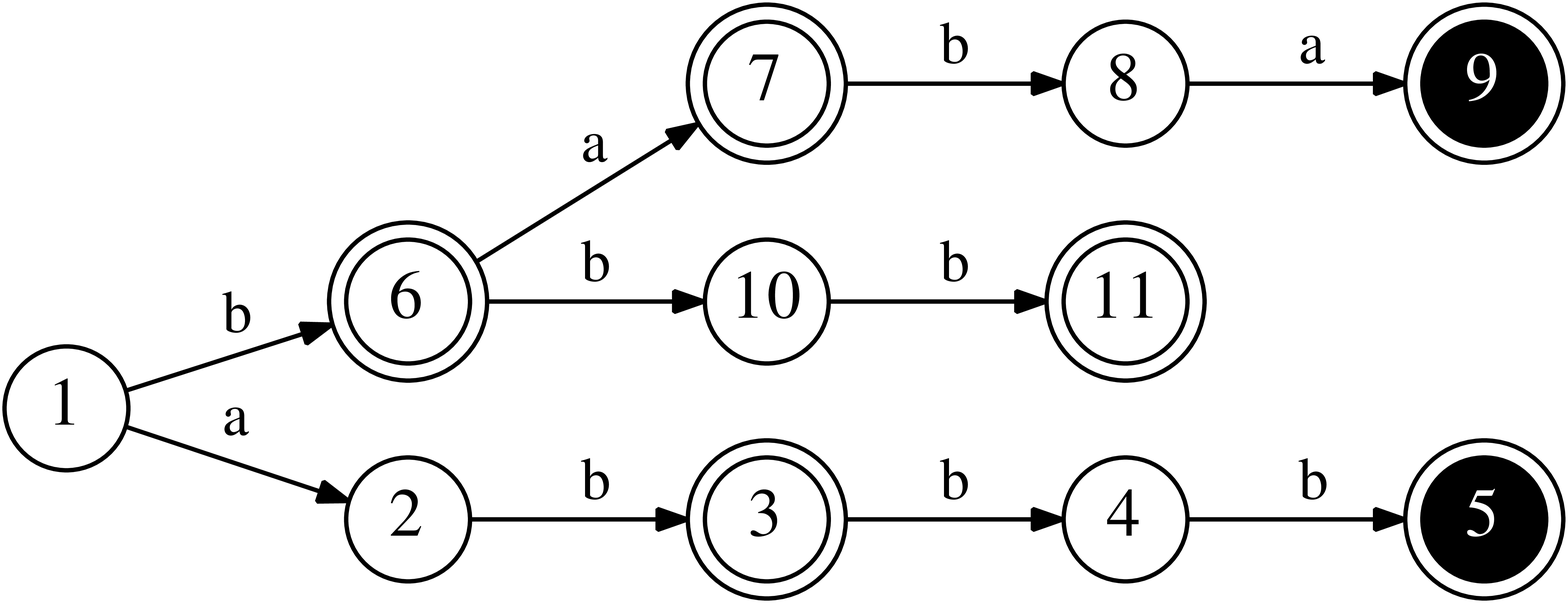}
  }
  \hfill
  \subfloat[The consistency graph for the APTA from Fig.~\ref{APTA}\label{con-graph}]{%
    \includegraphics[width=2in]{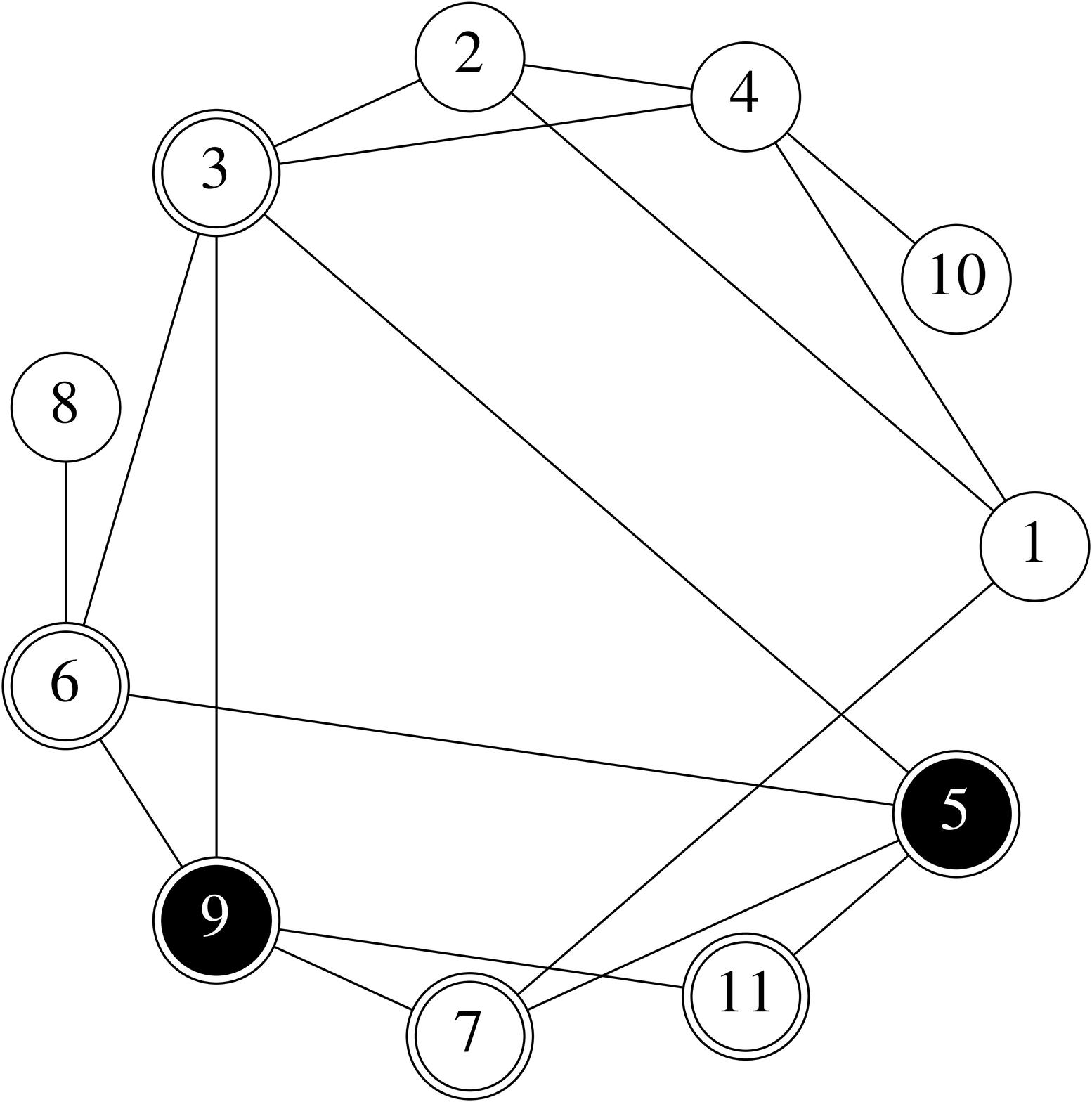}
  }
  \caption{An example of an APTA and its consistency graph}
  \label{apta+con}

\end{figure}

In~\cite{heule2010} Heule and Verwer proposed a compact translation of the DFA identification problem into SAT.
Here we briefly review the proposed technique, since our symmetry breaking predicates supplement it.
The first step of both the state-merging and the translation-to-SAT techniques 
is the augmented prefix tree acceptor (APTA) construction from the given examples $S_+$ and $S_-$.
An APTA is a tree-shaped automaton such that paths corresponding to two strings reach the same state $v$
if and only if these strings share the same prefix in which the last symbol corresponds to $v$.
We denote by $V$ the set of all APTA states; by $v_r$ the APTA root; by $V_+$ the set of accepting states;
and by $V_-$ the set of rejecting states.
Moreover, for state $v$ (except $v_r$) we denote its incoming symbol by $l(v)$ and its parent by $p(v)$.
The APTA for $S_+$ and $S_-$ mentioned above is shown in Fig.~\ref{APTA}.

The second step of the technique proposed in~\cite{heule2010} is the 
construction of the \textit{consistency graph} (CG) for the obtained APTA.
The set of nodes of the CG is identical to the set of APTA states.
Two CG nodes $v$ and $w$ are connected with an edge~(and called inconsistent) if
merging $v$ and $w$ in the APTA results in an inconsistency: an accepting state is merged with a rejecting state.
Let $E$ denote the set of CG edges.
The CG for APTA of Fig.~\ref{APTA} is shown in Fig.~\ref{con-graph}.

The key part of the algorithm is translating the DFA identification problem 
into a Boolean formula in conjunctive normal form (CNF) and using a SAT solver to find a satisfying assignment.
For a given set of examples and fixed DFA size $C$ the solver returns a satisfying assignment 
(that defines a DFA with $C$ states that is compliant with $S_+$ and $S_-$) or a message that it does not exist.
The main idea of this translation is to use a distinct color for every state of the 
identified DFA and to find a consistent mapping of APTA states to colors.
Three types of variables were used in the proposed compact translation:
\begin{enumerate}
  \item \textit{color} variables $x_{v,i} \equiv 1$ ($v \in V$; $1 \leq i \leq C$) if and only if APTA state $v$ has color $i$ 
  (APTA state $v$ will be merged into DFA state $i$);
  \item \textit{parent relation} variables $y_{l,i,j} \equiv 1$ ($l \in \Sigma$; $1 \leq i, j \leq C$) 
        if and only if DFA transition with symbol $l$ from state $i$ ends in state~$j$;
  \item \textit{accepting color} variables $z_i \equiv 1$ ($1 \leq i \leq C$) if and only if DFA state $i$ is accepting.
\end{enumerate}

Direct encoding, described in~\cite{heule2010}, uses only variables $x_{v,i}$; variables $y_{l,i,j}$ 
and $z_i$ are auxiliary and used in compact encoding predicates, which are described below.

The compact translation proposed in~\cite{heule2010} uses nine types of clauses:
\begin{enumerate}
  \item $x_{v,i} \Rightarrow z_{i}$ $(v \in V_+$; $1 \leq i \leq C)$~--
        definitions of $z_i$ values for accepting states $(\neg x_{v,i} \vee z_{i})$;
  \item $x_{v,i} \Rightarrow \neg z_{i}$ $(v \in V_-$; $1 \leq i \leq C)$~--
        definitions of $z_i$ values for rejecting states $(\neg x_{v,i} \vee \neg z_{i})$;
  \item $x_{v,1} \vee x_{v,2} \vee \ldots \vee x_{v,C}$ $(v \in V)$~-- 
        each state $v$ has at least one color;
  \item $x_{p(v),i} \wedge x_{v, j} \Rightarrow y_{l(v),i,j}$ $(v \in V \setminus \{v_r\}$; $1 \leq i,j \leq C)$~--
        a DFA transition is set when a state and its parent are colored $(y_{l(v),i,j} \vee \neg x_{p(v),i} \vee \neg x_{v, j})$;
  \item $y_{l,i,j} \Rightarrow \neg y_{l,i,k}$ $(l \in \Sigma$; $1 \leq i,j,k \leq C$; $j < k)$~--
        each DFA transition can target at most one state $(\neg y_{l,i,j} \vee \neg y_{l,i,k})$;
  \item $\neg x_{v,i} \vee \neg x_{v,j}$ $(v \in V$; $1 \leq i < j \leq C)$~--
        each state has at most one color;
  \item $y_{l, i, 1} \vee y_{l, i, 2} \vee \ldots \vee y_{l, i, C}$ $(l \in \Sigma$; $1 \leq i \leq C)$~--
        each DFA transition must target at least one state;
  \item $y_{l(v),i,j} \wedge x_{p(v),i} \Rightarrow x_{v, j}$ $(v \in V \setminus \{v_r\}$; $1 \leq i,j \leq C)$~--
        state color is set when DFA transition and parent color are set $(\neg y_{l(v),i,j} \vee \neg x_{p(v),i} \vee x_{v, j})$;
  \item $x_{v,i} \Rightarrow \neg x_{w,i}$ $((v,w) \in E$; $1 \leq i \leq C)$~--
        the colors of two states connected with an edge in the consistency graph must be different $(\neg x_{v,i} \vee \neg x_{w,i})$.
\end{enumerate}

Thus, the constructed formula consists of $\mathcal{O}(C^2 |V|)$ clauses and, 
if the SAT solver finds a solution, we can identify the DFA.

To find a minimal DFA, the authors use incremental SAT solving.
The initial DFA size $C$ is equal to the size of a \textit{large clique} found in the CG.
To find such a clique, a greedy algorithm proposed in~\cite{heule2010} can be applied.
Then the minimal DFA is found by iterating over the DFA size $C$ until the formula is satisfied.
Algorithm~\ref{exact_algo} illustrates this approach.

The found clique was also used to perform symmetry breaking: in any valid coloring of a graph, 
all states in a clique must have a different color.
Thus, we can fix the state colors in the clique during the preprocessing step.
Later we will see that the max-clique symmetry breaking is not compatible with the one proposed in this paper.

To reduce the SAT search space significantly, the authors applied several EDSM steps before translating the problem to SAT.
Since EDSM cannot guarantee the minimality of solution, we will omit the consideration of this step in our paper.

\begin{algorithm}[ht]
 \DontPrintSemicolon
 \SetKwData{SS}{SS}\SetKwData{APTA}{APTA}\SetKwData{CG}{CG}\SetKwData{minSize}{minSize}
 \SetKwData{maxSize}{maxSize}\SetKwData{clique}{clique}\SetKwData{solver}{solver}\SetKwData{TL}{TL}
 \SetKwData{dimacsFile}{dimacsFile}\SetKwData{SBStrategy}{SBStrategy}\SetKwData{colors}{colors}
 \SetKwData{result}{result}\SetKwData{solution}{solution}\SetKwData{DFA}{DFA}\SetKwData{size}{size}
 \SetKwFunction{buildAPTA}{buildAPTA}\SetKwFunction{buildCG}{buildCG}
 \SetKwFunction{findClique}{findClique}\SetKwFunction{max}{max}
 \SetKwFunction{generateFileInDimacsFormat}{generateFileInDimacsFormat}
 \SetKwFunction{buildSolution}{buildSolution}\SetKwData{getSolution}{getSolution}
 \SetKwFunction{solve}{solve}

  \KwData{set of strings \SS, minimum size of the target DFA \minSize, maximum size of the target DFA \maxSize, 
  symmetry breaking strategy \SBStrategy, external SAT-solver \solver, time limit for SAT-solver \TL}
  \APTA $\leftarrow$ \buildAPTA{\SS}\\
  \CG $\leftarrow$ \buildCG{\APTA}\\
  \clique $\leftarrow$ \findClique{\APTA, \CG}\\
  \minSize $\leftarrow$ \max{\minSize, \clique.\size}\\
  \For{\colors $\leftarrow$ \minSize \KwTo \maxSize}{
   \dimacsFile $\leftarrow$ \generateFileInDimacsFormat{\APTA, \CG, \colors, \SBStrategy}\\
   \result $\leftarrow$ \solver.\solve{\dimacsFile, \TL}\\
   \If{\upshape \result is `SAT'}{
    \solution $\leftarrow$ \solver.\getSolution{}\\
    \DFA $\leftarrow$ \buildSolution{\solution}\\
    \KwRet{\DFA}
   }
  }
  \KwRet{\upshape null}

 \caption{Scheme of the SAT-based algorithm}
\label{exact_algo}
\end{algorithm}

\section{Learning DFA from Noisy Samples}
  \label{noisy}

The translation described in the previous section deals with exact DFA identification.
In this section we show how to modify the translation in order to apply it to noisy examples.
We assume that not more than $K$ attribution labels of the given training strings were randomly flipped.
Solving this problem was the goal of the GECCO 2004 Noisy DFA competition~\cite{gecco2004}
(with $K$ equal to $10$ percent of the number of the given training strings).
The EA with smart state labeling was later proposed in~\cite{lucas2005},
and since that time it has been, to the best of our knowledge, 
the state-of-the-art technique for learning DFA from noisy training data.

In the noisy case we cannot use the APTA node consistency:
we cannot determine whether an accepting state is merged with a rejecting state since 
correct string labels are unknown.
Thus we cannot use CG and the max-clique symmetry breaking.

The idea of our modification is rather simple: for each labeled state of the APTA we 
define a variable which states whether the label can be flipped.
The number of flips is limited by $K$.
Formally, for each $v \in V_{\pm} = V_+ \cup V_-$ we define $f_v$ which is true if 
and only if the label of state $v$ can (but does not have to) be incorrect (\textbf{f}lipped).
Using these variables, we can modify the translation proposed in~\cite{heule2010} to take into account mistakes in string labels.
To do this, we change the $z_i$ definition clauses (items 1 and 2 from list in Section~\ref{exact-DFA}):
because of mistake possibility they hold in case $f_v$ is false.
Thus, new $z_i$ value definitions are expressed in the following way:
$\neg f_v \Rightarrow (x_{v, i} \Rightarrow z_i)$ for $v \in V_+$; 
$\neg f_v \Rightarrow (x_{v, i} \Rightarrow \neg z_i)$ for $v \in V_-$.

To limit the number of corrections to $K$ we use an 
auxiliary array of $K$ integer variables. 
This array stores the numbers of the APTA states for which labels can be flipped.
Thus, $f_v$ is true if and only if the array contains $v$.
To avoid the consideration of isomorphic permutations we enforce the array to be sorted in the increasing order.

To represent the auxiliary array as a Boolean formula we define variables $r_{i, v}$ 
for $1 \leq i \leq K$ and $v \in V_{\pm} = \{v_1, \ldots, v_W\}$.
$r_{i,v}$ is true if and only if $v$ is stored in the $i$-th position of the array. 
To connect variables $f_v$ with $r_{i,v}$ we add so-called channeling constrains:
$f_v \Leftrightarrow (r_{1, v} \vee \ldots \vee r_{K, v})$ for each $v \in V_{\pm}$.

We have to state that exactly one $r_{i, v}$ is true for each position $i$ in the auxiliary array.
To achieve that we use the order encoding method~\cite{barahona2014}. 
We add auxiliary \textbf{o}rder variables $o_{i, v}$ for $1 \leq i \leq K$ and $v \in V_{\pm}  = \{v_1, \ldots, v_W \}$.
We assume that $o_{i, v}$ for $v \in \{v_1, \ldots,  v_j\}$ and $\neg o_{i, v}$ for $v \in \{v_{j+1}, \ldots,  v_W\}$ for some $j$.
This can be expressed by the following constraint: $o_{i, v_{j+1}} \Rightarrow o_{i, v_j}$ for $1 \leq j < W$.
Now we define that $r_{i, v_{j}} \Leftrightarrow o_{i, v_{j}} \wedge \neg o_{i, v_{j+1}}$.
We also add clauses $o_{i, v_j} \Rightarrow o_{i+1, v_{j+1}}$ (for $1 \leq i < K$ and $1 \leq j < W$) 
to store corrections in the increasing order.

The proposed constraints in CNF are listed in Table~\ref{noisy-table}; 
there are $\mathcal{O}(C |V_\pm| + K |V_\pm|)$ clauses.
Thus, to modify the translation for the noiseless case to deal with noise we can replace 
the $z_i$ value definition and inconsistency clauses 
(items 1, 2 and 9 from list in Section~\ref{exact-DFA})
with the ones listed in Table~\ref{noisy-table}.

\begin{table}
\centering
\caption{Clauses for noisy DFA identification}
\begin{tabular}{lll}
  Clauses & CNF representation & Range \\
  \hline
  $\neg f_v \Rightarrow (x_{v, i} \Rightarrow z_i)$ & 
  $\neg x_{v, j} \vee z_j \vee f_v$ & 
  $1 \leq j \leq C$; $v \in V_+$\\

  $\neg f_v \Rightarrow (x_{v, i} \Rightarrow \neg z_i)$ &
  $\neg x_{v, j} \vee \neg z_j \vee f_v$ & 
  $1 \leq j \leq C$; $v \in V_-$\\

  \hline
  $f_v \Rightarrow (r_{1, v} \vee \ldots \vee r_{K, v})$ &
  $\neg f_v \vee r_{1, v} \vee \ldots \vee r_{K, v}$ & 
  $v \in V_{\pm}$\\

  $r_{i, v} \Rightarrow f_v$ &
  $\neg r_{i, v} \vee f_v$ & 
  $1 \leq i \leq K$; $v \in V_{\pm}$\\
  \hline
  
  $r_{i, v_j} \Rightarrow o_{i, v_j}$ &
  $\neg r_{i, v_j} \vee o_{i, v_j}$ & 
  $1 \leq i \leq K$; $1 \leq j \leq W$\\

  $r_{i, v_j} \Rightarrow \neg o_{i, v_{j+1}}$ &
  $\neg r_{i, v_j} \vee \neg o_{i, v_{j+1}}$ & 
  $1 \leq i \leq K$; $1 \leq j < W$\\

  $o_{i, v_j} \wedge \neg o_{i, v_{j+1}} \Rightarrow r_{i, v_j}$ &
  $\neg o_{i,v_j} \vee o_{i, v_{j+1}} \vee r_{i, v_j}$ & 
  $1 \leq i \leq K$; $1 \leq j < W$\\

  $o_{K, v_W} \Rightarrow r_{K, v_W}$ &
  $\neg o_{K,v_W} \vee r_{K, v_W}$ & 
  \\

\hline

  $o_{i, v_{j+1}} \Rightarrow o_{i, v_j}$ &
  $\neg o_{i, v_{j+1}} \vee o_{i, v_j}$ &
  $1 \leq i \leq K$; $1 \leq j < W$\\

  $o_{i, v_j} \Rightarrow o_{i+1, v_{j+1}}$ &
  $\neg o_{i, v_j} \vee o_{i+1, v_{j+1}}$ &
  $1 \leq i < K$; $1 \leq j < W$\\
  

  \hline
\end{tabular}
\label{noisy-table}
\end{table}

\section{DFS-based Symmetry Breaking Predicates}
  \label{dfs-sbp}

In this section we propose a way to fix automata state enumeration to avoid consideration of isomorphic automata during SAT solving.
The main idea of our symmetry breaking is to enforce DFA states to be enumerated in the depth-first search (DFS) order.
Thus only one representative of each equivalence class with respect to the isomorphic relation will be processed.

\begin{figure}[!ht]
  \subfloat[DFS-enumerated DFA with bolded DFS tree edges\label{dfs:dfa}]{%
    \includegraphics[width=2.2in]{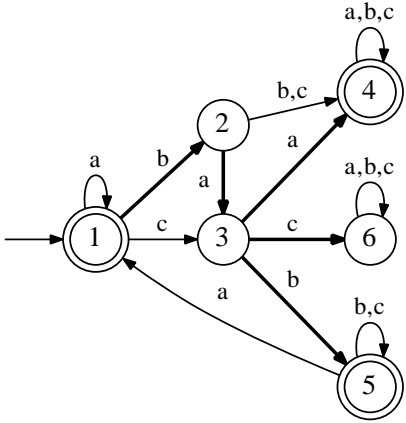}
  }
  \hfill
  \subfloat[DFS auxiliary array. The elements correspond to DFA states, and the transitions are the ones used in the DFS traversal\label{dfs:array}]{%
    \includegraphics[width=2.5in]{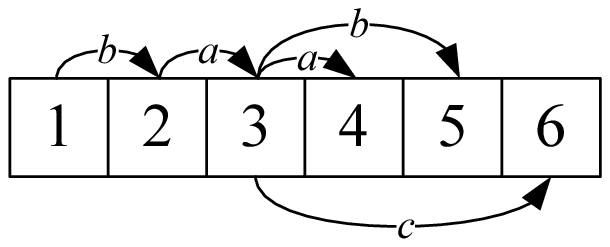}
  }
  \caption{An example of a DFS-numerated DFA and its DFS auxiliary array}
  \label{dfs}

\end{figure}

It is necessary to find all adjacent unvisited states for each unvisited state of the DFA during DFS processing.
First the algorithm handles the start DFA state. Then the algorithm
processes the children of this state and recursively runs for each of them.
We will use an auxiliary structure, an array with transitions which connect its elements.
These elements are the numbers of states ordered according to the DFS enumeration. 
Each transition connecting the elements of the array will be a copy of a DFA transition which was used by DFS.
Since our transitions are labeled with symbols from $\Sigma$, 
we process child states in the alphabetical order of symbols $l$ on transitions $i \xrightarrow{l} j$.
We call a DFA \textit{DFS-enumerated} if its auxiliary array is filled by consecutive numbers in the increasing
order starting from $1$. An example of a DFS-enumerated DFA with six states is shown in Fig.~\ref{dfs:dfa} 
(DFS tree transitions that were used to add states into the array are marked bold);
the DFS auxiliary array for this DFA is shown in Fig.~\ref{bfs:array}.
The DFA shown in Fig.~\ref{dfa-example} is not DFS-enumerated~-- DFS first handles state $3$ rather than state $2$ 
(we consider $1 \xrightarrow{a} 3$ before $1 \xrightarrow{b} 2$).

We propose the constraints that enforce DFA to be DFS-enumerated.
We assume that translation of a given DFA identification problem to SAT deals with Boolean variables 
$y_{l,i,j}$ ($l \in \Sigma$; $1 \leq i, j \leq C$) 
to set the DFA transition function: $y_{l,i,j} \equiv 1$
if and only if the transition with symbol $l$ from state $i$ ends in state $j$.

The main idea is to determine each state's parent in the DFS-tree and set constrains between states' parents.
We store \textbf{p}arents in values $p_{j,i}$ (for each $1 \leq i < j \leq C$).
$p_{j,i}$ is true if and only if state $i$ is the parent of $j$ in the DFS-tree.
Each state except the initial one must have a parent with a smaller number, thus
$$\bigwedge\limits_{2 \leq j \leq C} (p_{j,1} \vee p_{j,2} \vee \ldots \vee p_{j,j-1}).$$ 

We set parent variables $p_{j,i}$ through $y_{l,i,j}$ using auxiliary variables $t_{i,j}$.
In the DFS-enumeration state $j$ was added into the array while processing the state with maximal number $i$ among states that have a transition to $j$: 
$$\bigwedge\limits_{1 \leq i < j \leq C} 
(p_{j,i} \Leftrightarrow t_{i,j} \wedge \neg t_{i+1,j} \wedge \ldots \wedge \neg t_{j-1,j}),$$
where $t_{i,j} \equiv 1$ if and only if there is a \textbf{t}ransition between $i$ and $j$.
We define these auxiliary variables using $y_{l,i,j}$:
$$\bigwedge\limits_{1 \leq i < j \leq C} (t_{i,j} \Leftrightarrow y_{l_1,i,j} \vee \ldots \vee y_{l_L,i,j}).$$

\begin{figure}
  \centering%
  \includegraphics[width=1.9in]{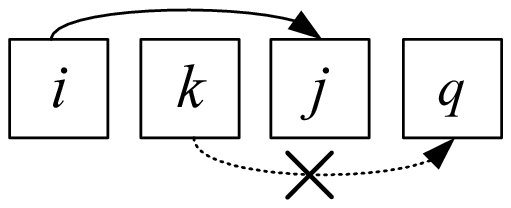}
  \caption{Part of the array illustrating the parent ordering predicates. 
           The transitions show parent relations. The dotted transition is not allowed due to the DFS-enumeration}
  \label{dfsa1}
\end{figure}

Moreover, in the DFS-enumeration states' parents must be ordered. 
If $i$ is the parent of the state $j$ and $k$ is a state between $i$ and $j$
($i < k < j$) then there is no transition from state $k$ to state $q$ which
is bigger then $j$ (see Fig.~\ref{dfsa1}):
$$\bigwedge\limits_{1 \leq i < k < j < q < C} (p_{j,i} \Rightarrow \neg t_{k,q}).$$

Now to enforce the DFA to be DFS-enumerated we have to order children in the alphabetical order of symbols on transitions.
We consider two cases: alphabet $\Sigma$ consists of two symbols $\{a,b\}$ and more than two symbols $\{l_1, \ldots, l_L\}$.
In the case of two symbols only two states $j$ and $k$ can have the same parent $i$ (where without loss of generality $j < k$).
In this case we force the transition that starts in state $i$ labeled with symbol $a$ to end in state $j$ instead of $k$:
$$\bigwedge\limits_{1 \leq i < j < k < C}
  (p_{j,i} \wedge p_{k,i} \Rightarrow y_{a,i,j}).$$

In the second case we have to introduce the third type of variables in our symmetry breaking predicates.
We store the alphabetically \textbf{m}inimal symbol on transitions between states: 
$m_{l,i,j}$ is true if and only if there is a transition $i \xrightarrow{l} j$ and 
there is no such transition with an alphabetically smaller symbol.
We connect these variables with DFA transitions by adding the following channeling predicates:
$$\bigwedge\limits_{1 \leq i < j \leq C} \bigwedge\limits_{1 \leq n \leq L}
  (m_{l_n,i,j} \Leftrightarrow y_{l_n,i,j} \wedge \neg y_{l_{n-1},i,j} \wedge \ldots \wedge \neg y_{l_1,i,j}).$$

\begin{figure}
  \centering%
  \includegraphics[width=1.4in]{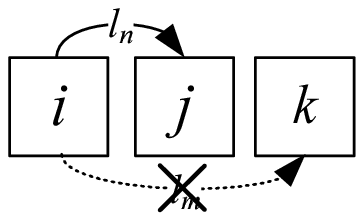}
  \caption{Illustration of alphabetical ordering predicates. 
           If $i$ is the parent of $j$ and $k$, $l_n$ ($l_m$) is the alphabetically minimal symbol on 
           transitions between $i$ and $j$ ($i$ and $k$) then $l_m$ cannot be alphabetically smaller than $l_n$}
  \label{dfsa2}
\end{figure}

Now it remains to arrange states $j$ and $k$ with the same parent $i$
in the alphabetical order of minimal symbols on transitions between them and $i$ (see Fig.~\ref{dfsa2}):
$$\bigwedge\limits_{1 \leq i < j < k \leq C} \bigwedge\limits_{1 \leq m < n \leq L}
  (p_{j,i} \wedge p_{k,i} \wedge m_{l_n,i,j} \Rightarrow \neg m_{l_m,i,k}).$$


Thus we propose symmetry breaking predicates that are composed of the listed constraints.
Predicates (for the case of three or more symbols) translated into $\mathcal{O}(C^4 + C^3 L^2)$ CNF clauses are listed in Table~\ref{sbp-table}.

\section{BFS-based Symmetry Breaking Predicates}
  \label{bfs-sbp}

In this section we consider a modification of the DFS-based approach which
enforce DFA states to be enumerated in the breadth-first search (BFS) order.
This idea was also used in function \texttt{NatOrder} in the state-merging approach from~\cite{lambeau2008} 
and the Move To Front reorganization algorithm used in the genetic algorithm~\cite{chambers2010}.

\begin{figure}[!ht]
  \subfloat[BFS-enumerated DFA with bolded BFS-tree edges\label{bfs:dfa}]{%
    \includegraphics[width=2.2in]{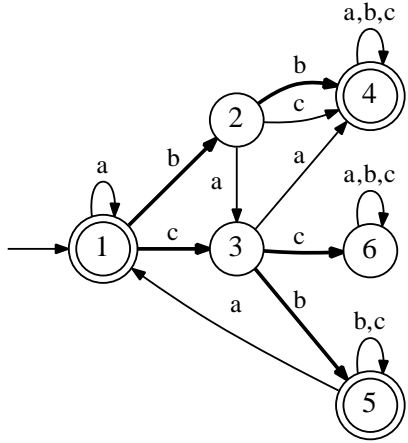}
  }
  \hfill
  \subfloat[BFS queue. Cells correspond to DFA states, transitions correspond to enqueues\label{bfs:array}]{%
    \includegraphics[width=2.5in]{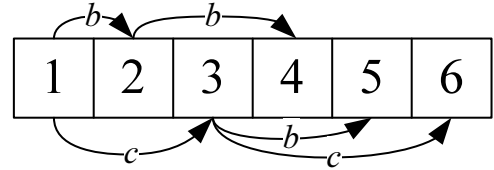}
  }
  \caption{An example of BFS-enumerated DFA and its BFS queue}
  \label{bfs}

\end{figure}

BFS uses the \textit{queue} data structure to store intermediate results as it traverses the graph.
First we enqueue the initial DFA state.
While the queue is not empty we deque a state $i$ and enqueue any direct 
child states $j$ that have not yet been discovered (enqueued before).
We enqueue child states in alphabetical order of symbols $l$ on transitions $i \xrightarrow{l} j$ the same as in the DFS-based approach.
We call a DFA \textit{BFS-enumerated} if its states are enumerated in dequeuing (equals to enqueuing) order.
An example of a BFS-enumerated DFA with six states is shown in Fig.~\ref{bfs:dfa} 
(BFS tree transitions that were used to enqueue states are marked bold);
BFS enqueues are shown in Fig.~\ref{bfs:array}.
The DFA shown in Fig.~\ref{dfa-example} is not BFS-enumerated~-- BFS first dequeues state $3$ rather than state $2$ 
(we consider $1 \xrightarrow{a} 3$ before $1 \xrightarrow{b} 2$).

All variables which were used for the DFS enumeration are also used for the BFS enumeration, but 
we have to consider the constraints which must be changed.
In the BFS-enumeration $p_{j,i}$ variable definition is changed:
state $j$ should be enqueued while processing the state with the minimal number $i$ among the states that have a transition to $j$: 
$$\bigwedge\limits_{1 \leq i < j \leq C} 
(p_{j,i} \Leftrightarrow t_{i,j} \wedge \neg t_{i-1,j} \wedge \ldots \wedge \neg t_{1,j}).$$

\begin{figure}
  \centering%
  \includegraphics[width=1.9in]{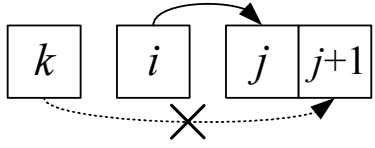}
  \caption{Part of the queue illustrating the parent ordering predicates. 
           Transitions show parent relations. The dotted transition is not allowed due to BFS-enumeration}
  \label{a1}
\end{figure}

In the BFS-enumeration states' parents ordering constraints are also changed:
state $j$ must be enqueued before the next state $j+1$, thus the next 
state's parent $k$ cannot be less than the current state's parent $i$ (see Fig.~\ref{a1}):
$$\bigwedge\limits_{1 \leq k < i < j < C} (p_{j,i} \Rightarrow \neg p_{j+1,k}).$$

\begin{figure}
  \centering%
  \includegraphics[width=1.4in]{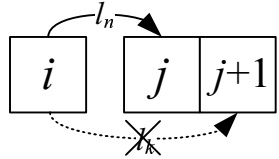}
  \caption{Illustration of alphabetical ordering predicates. 
           If $i$ is the parent of $j$ and $j+1$, $l_n$ ($l_k$) is the alphabetically minimal symbol on 
           transitions between $i$ and $j$ ($i$ and $j+1$) then $l_k$ cannot be alphabetically smaller than $l_n$}
  \label{a2}
\end{figure}

It is enough to consider only two consecutive states in the constraints where two states with the same parent were considered (see Fig.~\ref{a2} for the second type of the constraints):
$$\bigwedge\limits_{1 \leq i < j < C}
  (p_{j,i} \wedge p_{j+1,i} \Rightarrow y_{a,i,j}),$$
$$\bigwedge\limits_{1 \leq i < j < C} \bigwedge\limits_{1 \leq m < n \leq L}
  (p_{j,i} \wedge p_{j+1,i} \wedge m_{l_n,i,j} \Rightarrow \neg m_{l_m,i,j+1}).$$

Predicates (for the case of three or more symbols) translated into $\mathcal{O}(C^3 + C^2 L^2)$ CNF clauses are listed in Table~\ref{sbp-table}.
Our implementation of the proposed predicates and all algorithms can be found on the github repository of our laboratory 
\footnote{\url{https://github.com/ctlab/DFA-Inductor}}.

\begin{table}
\centering
\caption{DFS-based and BFS-based symmetry breaking clauses}
\scalebox{0.865}{
\begin{tabular}{llll}

   & Clauses & CNF representation & Range \\
  \hline
  \multirow{7}{*}{\rotatebox{90}{Both}}&
  $t_{i,j} \Rightarrow (y_{l_1,i,j} \vee \ldots \vee y_{l_L,i,j})$ &
  $\neg t_{i,j} \vee y_{l_1,i,j} \vee \ldots \vee y_{l_L,i,j}$ & 
  $1 \leq i < j \leq C$
  \\
  &
  $y_{i,j,l} \Rightarrow t_{i,j}$ &
  $\neg y_{l,i,j} \vee t_{i,j}$ & 
  $1 \leq i < j \leq C$; $l \in \Sigma$ 
  \\
  &
  $p_{j,i} \Rightarrow t_{i,j}$ &
  $\neg p_{j,i} \vee t_{i,j}$ & 
  $1 \leq i < j \leq C$ 
  \\
  &
  $p_{j,1} \vee p_{j,2} \vee \ldots \vee p_{j,j-1}$ &
  $p_{j,1} \vee p_{j,2} \vee \ldots \vee p_{j,j-1}$ & 
  $2 \leq j \leq C$ 
  \\
  &
  $m_{l,i,j} \Rightarrow y_{l,i,j}$ &
  $\neg m_{l,i,j} \vee y_{l,i,j}$ & 
  $1 \leq i < j \leq C$; $l \in \Sigma$ 
  \\
  &
  $m_{l_n,i,j} \Rightarrow \neg y_{l_k,i,j}$ &
  $\neg m_{l_n,i,j} \vee \neg y_{l_k,i,j}$ & 
  $1 \leq i < j \leq C$; $1 \leq k < n \leq L$
  \\
  &
  \vtop{\hbox{\strut $(y_{l_n,i,j} \wedge \neg y_{l_{n-1},i,j} \wedge \ldots$}
        \hbox{\strut \qquad $\neg y_{l_1,i,j}) \Rightarrow m_{l_n,i,j}$}} &
  \vtop{\hbox{\strut $\neg y_{l_n,i,j} \vee y_{l_{n-1},i,j} \vee \ldots$}
        \hbox{\strut \qquad $\vee y_{l_1,i,j} \vee m_{l_n,i,j}$}} &

  $1 \leq i < j \leq C$; $1 \leq n \leq L$ 
  \\  
  \hline
  \multirow{5}{*}{\rotatebox{90}{DFS}}&
  $p_{j,i} \Rightarrow \neg t_{k,j}$ &
  $\neg p_{j,i} \vee \neg t_{k,j}$ & 
  $1 \leq i < k < j \leq C$
  \\
  &
  $(t_{i,j} \wedge \neg t_{i+1,j} \wedge \ldots \wedge \neg t_{j-1,j}) \Rightarrow p_{j,i}$ &
  $\neg t_{i,j} \vee t_{i+1,j} \vee \ldots \vee t_{j-1,j} \vee p_{j,i}$ &
  $1 \leq i < j \leq C$
  \\
  &
  $p_{j,i} \Rightarrow \neg t_{k,q}$ & 
  $\neg p_{j,i} \vee \neg t_{k,q}$ & 
  $1 \leq i < k < j < q \leq C$
  \\
  &
  $(p_{j,i} \wedge p_{k,i} \wedge m_{l_n,i,j}) \Rightarrow \neg m_{l_m,i,k}$ &
  $\neg p_{j,i} \vee \neg p_{k,i} \vee \neg m_{l_n,i,j} \vee \neg m_{l_m,i,k}$ &
  $1 \leq i < j < k \leq C$; $1 \leq m < n \leq L$
  \\
  \hline
  \multirow{5}{*}{\rotatebox{90}{BFS}}&
  $p_{j,i} \Rightarrow \neg t_{k,j}$ &
  $\neg p_{j,i} \vee \neg t_{k,j}$ & 
  $1 \leq k < i < j \leq C$
  \\
  &
  $(t_{i,j} \wedge \neg t_{i-1,j} \wedge \ldots \wedge \neg t_{1,j}) \Rightarrow p_{j,i}$ &
  $\neg t_{i,j} \vee t_{i-1,j} \vee \ldots \vee t_{1,j} \vee p_{j,i}$ &
  $1 \leq i < j \leq C$
  \\
  &
  $p_{j,i} \Rightarrow \neg p_{j+1,k}$ & 
  $\neg p_{j,i} \vee \neg p_{j+1,k}$ & 
  $1 \leq k < i < j < C$
  \\
  &
  $(p_{j,i} \wedge p_{j+1,i} \wedge m_{l_n,i,j}) \Rightarrow \neg m_{l_m,i,j+1}$ &
  $\neg p_{j,i} \vee \neg p_{j+1,i} \vee \neg m_{l_n,i,j} \vee \neg m_{l_m,i,j+1}$ &
  $1 \leq i < j < C$; $1 \leq m < n \leq L$
  \\
  \hline
\end{tabular}
}
\label{sbp-table}
\end{table}

\section{The find-all problem}
In this section we pay our attention to the problem of finding all non-isomorphic DFA with the minimal number of states which 
are consistent with a given set of strings. We propose the way to modify the SAT-based method in order to apply it to this problem. 
We consider two ways of using SAT-solvers: restarting a non-incremental solver after finding each automaton and using an incremental solver:
if such a solver finds a solution, it retains its state and is ready to accept new clauses. 
The most common interface and technique for incremental SAT-solving was proposed in~\cite{een2004}.
We also consider the heuristic Backtracking method as a baseline for comparing it with SAT-based ones.

\subsection{SAT-based methods}

The main idea of SAT-based methods is to ban satisfying interpretations which have already been found.
It is obvious that if the proposed symmetry breaking predicates are not used then this approach
finds a lot of isomorphic automata (exactly $n!$ for each equivalence class where $n$ is the DFA size). 
Since max-clique predicates fix $k$ colors only (where $k$ is the clique size), 
the proposed algorithm finds $(n-k)!$ isomorphic automatons which is also bad.
The BFS-based and DFS-based symmetry breaking predicates allow to ban isomorphic DFA from 
one equality class by banning accordingly enumerated representative.
It is easy to implement by adding an additional clause into the Boolean formula. 
Since we know that $y_{l,i,j}$ variables define the target DFA entirely, it is enough 
to forbid only values of these variables from found interpretation:
$$\neg y_{1} \vee \neg y_{2} \vee \ldots \vee \neg y_{n|\Sigma|},$$
where $y_{k}$ is some true $y_{l,i,j}$ from the found interpretation for $1 < k < n|\Sigma|$.

There are two different ways of using SAT-solvers as it was stated above.
First, we can restart a non-incremental SAT-solver with the new Boolean formula with the 
additional clause after finding each automaton.
The second approach is based on an incremental SAT-solver: 
after each found automaton we add the additional clause to the solver and continue its execution.

\begin{figure}[!ht]
  \subfloat[An example of an APTA for $S_+=\{ab, b, ba, bbb\}$ and $S_-=\{abbb\}$\label{APTA-used}]{%
    \includegraphics[width=3in]{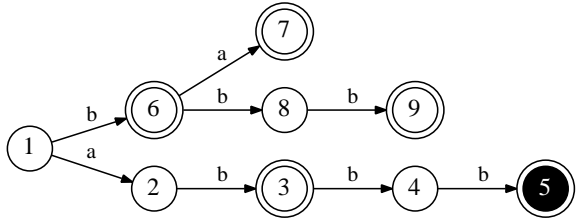}
  }
  \hfill
  \subfloat[The DFA is built by the APTA from Fig.~\ref{APTA-used} with unused $a$-labeled transition from the state $2$\label{DFA-used}]{%
    \includegraphics[width=2in]{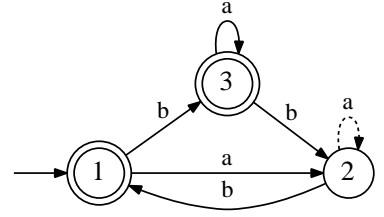}
  }
  \caption{An example of an APTA and its consistent DFA}
  \label{used}
\end{figure}

It is necessary to mention the case when some transitions of the found DFA are not
covered by the APTA. It means that there are some \textit{free} transitions which are not used during processing
any given word and each such transition can end in any state, since this does not influence the consistency
of the DFA with a given set of strings.
But in the case of the find-all problem we do not wish to find all these automatons differed
only by such transitions. Thus we propose the way to force all free transitions to finish in
the same state as they start. In other words we force them to be a loop. To achieve that we add 
auxiliary `\textbf{u}sed' variables: $u_{l,i}$ is true if and only if there is a $l$-labeled APTA edge from 
the $i$-colored state:
$$\bigwedge\limits_{l \in \Sigma} \bigwedge\limits_{1 \leq i \leq C} u_{l,i} \Leftrightarrow x_{1,i} \vee \ldots \vee x_{|V_l|,i},$$
where $V_l$ is the set of all the APTA states which have an outcoming edge labelled with $l$. 
To force unused transitions to be a loop we add the following constraints:
$$\bigwedge\limits_{l \in \Sigma} \bigwedge\limits_{1 \leq i \leq C} \neg u_{l,i} \Rightarrow y_{l,i,i}.$$
These additional constraints are translated into $\mathcal{O}(C |V|)$ clauses. See Fig.~\ref{used} for 
the example of an APTA for  $S_+=\{ab, b, ba, bbb\}$ and $S_-=\{abbb\}$ (it is the same example as on the Fig.~\ref{dfa-example} and Fig.~\ref{apta+con} but 
without the string $baba$ in the $S_-$) and its consistent DFA with unused transition. If
we add the proposed constraints then this transition will be forced to be a loop as shown by dashed line at the Fig.~\ref{DFA-used}.

\subsection{Backtracking method}

The solution based on backtracking does not use any external tools like SAT-solvers. 
This algorithm works as follows. Initially there is an empty DFA with $n$ states.
Also there is a \textit{frontier}~-- the set of edges from the APTA which are not represented yet in
the DFA. Initially the frontier contains all the outcoming edges of the APTA root. The recursive
function \texttt{Backtracking} maintains the frontier in the proper state.
If the frontier is not empty, then the function tries to augment the DFA with one of its edges.
Each found DFA is checked to be consistent with the APTA and if the DFA complies with it
then an updated frontier is found. If the frontier is empty then the DFA is checked for
completeness (a DFA is complete if there are transitions from each state labeled with all alphabet symbols). 
If it is not complete and there are nodes which have the number of outcoming edges less 
then the alphabet size then we add missing edges as loops with a function \texttt{MakeComplete}.
Algorithm~\ref{backtracking_algo} illustrates the solution. The function \texttt{FindNewFrontier}
returns the new frontier for the augmented DFA or null if the DFA is inconsistent with the APTA.

\begin{algorithm}[ht]
 \SetKwData{DFA}{DFA}\SetKwData{DFAPrime}{DFA$'$}\SetKwData{frontierPrime}{frontier$'$}\SetKwData{APTA}{APTA}\SetKwData{frontier}{frontier}\SetKwData{trans}{edge}
 \SetKwData{source}{source}\SetKwData{dest}{destination}\SetKwData{DFAset}{DFAset}\SetKwData{lab}{label}
 \SetKwFunction{MakeComplete}{MakeComplete}\SetKwFunction{FindNewFrontier}{FindNewFrontier}
 \SetKwFunction{Backtracking}{Backtracking}\SetKwFunction{add}{add}
 \KwData{augmented prefix tree acceptor \APTA, current \DFA (initially empty), \frontier (initially contains all APTA root outcoming edges)}
 \DFAset $\leftarrow$ new Set$<$DFA$>$\\
 \trans $\leftarrow$ any edge from \frontier\\
 \ForEach{\dest $\in 1..|S|$}{
  \source $\leftarrow$ the state of \DFA from which \trans should be added\\
  \DFAPrime $\leftarrow$ \DFA $\cup$ transition(\source, \dest, \trans.\lab)\\
  \frontierPrime $\leftarrow$ \FindNewFrontier{\APTA, \DFAPrime, \frontier}\\
  \If{\upshape \frontierPrime $\ne null$}{
   \uIf{\upshape \frontierPrime = $\varnothing$}{
    \DFAset.\add{\MakeComplete{\DFAPrime}}\\
   }\Else{
    \DFAset.\add{\Backtracking{\APTA, \DFAPrime, \frontierPrime}}
   }
  }
 }
 \KwRet{\upshape \DFAset}
 \caption{Backtracking solution}
\label{backtracking_algo}
\end{algorithm}

\section{Experiments}
  \label{experiments}

All experiments were performed using a machine with an AMD Opteron 6378 2.4 GHz processor running Ubuntu 14.04. 
All algorithms were implemented in Java, the \textit{lingeling} SAT-solver was used. Our own algorithm was used for generating problem instances 
for all the experiments based on randomly generated data sets. This algorithm builds a set of strings with
the following parameters: size $N$ of DFA to be generated, alphabet size $A$, the number $S$ of strings to be generated, 
noise level $K$ (percentage of attribution labels of generated strings which have to be randomly flipped). 

For exact DFA identification we used randomly generated instances. We used the following parameters: $N \in [10; 20]$, $A=2$, $S = 50 N$.
We compared the SAT-based approach with three types of symmetry breaking predicates: the max-clique algorithm from~\cite{heule2010} 
and the proposed DFS-based and BFS-based methods. Each experiment was repeated $100$ times. The time limit was set to 3600 seconds. 
Values in \textit{italics} mean that not all $100$ instances were solved within the time limit. If less than 
$50$ instances were solved then TL is shown instead of a value. The results are listed in Table~\ref{exact-results}.
It can be seen from the table that both DFS-based and BFS-based strategies clearly outperform the max-clique approach which is the current state-of-the-art.
BFS-based strategy in its turn outperforms DFS-based one when target automaton size is larger than $14$.

\begin{table}[ht]
\caption{Mean execution times in seconds of solving exact DFA identification}
\centering
\begin{tabular}{cccccc}
N & DFS              & & BFS   & & max-clique      \\
\hline
10 & 80.1            & & 80.3  & & 158.2           \\
11 & 109.7           & & 109.1 & & \textit{337.3}  \\
12 & 159.9           & & 151.6 & & \textit{684.4}  \\
13 & 200.5           & & 196.3 & & \textit{1146.3} \\
14 & 301.1           & & 254.4 & & TL              \\
15 & 406.4           & & 332.6 & & TL              \\
16 & 824.7           & & 560.4 & & TL              \\
17 & \textit{1217.6} & & 631.9 & & TL              \\
18 & \textit{1722.4} & & 685.7 & & TL              \\
19 & \textit{2294.1} & & 778.9 & & TL              \\
20 & TL              & & 903.1 & & TL              \\
\end{tabular}
\label{exact-results}
\end{table}

For noisy DFA identification we also used randomly generated instances. First we considered the case when the target DFA exists and the Boolean formula
is satisfiable. We used the following parameters: $N \in [5; 10]$, $A=2$, $S \in\{10 N, 25 N, 50 N\}$. 
We compared three methods: the SAT-based approach without any symmetry breaking predicates, our solution using BFS-based symmetry breaking predicates, 
and the current state-of-the-art EA from~\cite{lucas2005}. 
Each experiment was repeated $100$ times.
The time limit was set to $1800$ seconds. The initial experiments showed that the EA outperforms our method clearly when 
$K > 4 \%$. Therefore we set this parameter to $1\%-4\%$. We left only instances which were solved within the time limit. 
These results indicate that the BFS-based strategy finds the solution slightly faster than the current 
state-of-the-art EA only when $N$ is small ($<7$), noise level is small ($1\%-4\%$) and the number of strings is also small ($<50N$). 
But BFS-based strategy finds the solution extremely faster than SAT approach without the symmetry breaking strategy. 

The third experiment considered the case when the target DFA does not exist and the Boolean formula is unsatisfiable. Random dataset was also used here.
We tried to find the target DFA using the following parameters: $N \in [5; 7]$, $A = 2$,  $S = 50 N$, $K \in [1\%;2\%]$. 
The input set of strings was generated from an $(N+1)$-sized DFA. 
It should be noted that the EA from ~\cite{lucas2005} cannot determine that the automaton consistent with a given set of strings does not exist. 
On the other hand, all SAT-based methods are capable of that. Therefore we compared our implementation of compact SAT encoding without using symmetry breaking predicates (WO) 
and the same with the BFS-based predicates. Each experiment was repeated $100$ times and the time limit was set to $1800$ seconds again. 
The results are listed in Table~\ref{noisy-unsat-results}. It can be seen from the table that the BFS-based strategy reduces significantly the mean time of
determination that an automaton does not exist. 

\begin{table}[ht]
\caption{Mean execution times in seconds of solving noisy DFA identification when the target DFA does not exist}
\centering
\begin{tabular}{cccccc}
N & K & & BFS            & & WO              \\
\hline
5 & 1 & & 11.6           & & 257.1           \\
5 & 2 & & 46.4           & & \textit{1296.7} \\
6 & 1 & & 110.1          & & TL              \\
6 & 2 & & 581.7          & & TL              \\
7 & 1 & & \textit{995.3} & & TL              \\
7 & 2 & & TL             & & TL              \\
\end{tabular}
\label{noisy-unsat-results}
\end{table}

The last experiment concerned the find-all problem. A random dataset was also used here. We used the following parameters:
$N \in [5; 11]$, $A = 2$, $S \in\{5 N, 10 N, 25 N\}$. We compared the SAT-based method with the restarting strategy (REST), the SAT-based method with
the incremental strategy (INC) and the backtracking method (BTR). Each experiment was repeated $100$ times as well. The time limit was set to $3600$ seconds.
The first column in each subtable contains the number of instances which has more than one DFA in the solution ($>1$).
Values in \textit{italics} mean that not all $100$ instances were solved within the time limit. If less than 
$50$ instances were solved then TL is shown instead of a value. The results are listed in Table~\ref{find_all_table}.
It can be seen from the table SAT-based methods work significantly faster then backtracking one. 
As we see incremental strategy in its turn clearly outperforms restart strategy. It can be explained as incremental
SAT-solver saves its state but non-incremental solver does the same actions each execution.

\begin{table}
\centering
\caption{Mean execution times in seconds of SAT-based restart method, SAT-based incremental method and backtracking method}
\scalebox{0.94}{
\begin{tabular}{ccccccccccccccc}
\hline
\multirow{2}{*}{$|N|$} & \multicolumn{4}{c}{$S = 5|N|$}  & ~ & \multicolumn{4}{c}{$S = 10|N|$} & ~ & \multicolumn{4}{c}{$S = 25|N|$}\\\cline{2-5}\cline{7-10}\cline{12-15}
   & $>$1& REST   & INC   & BTR            & & $>$1& REST  & INC  & BTR             & & $>$1& REST & INC  & BTR            \\\hline
5  & 79  & 1.1    & 0.6   & 0.3            & & 31  & 1.0   & 0.8  & 0.4             & & 14  & 2.3  & 2.0  & 1.1           \\ 
6  & 77  & 1.9    & 0.9   & 0.9            & & 35  & 1.4   & 1.1  & 0.6             & & 20  & 4.6  & 3.4  & 2.0           \\ 
7  & 87  & 5.1    & 1.4   & 5.8            & & 25  & 2.3   & 1.6  & 2.6             & & 5   & 5.8  & 5.2  & 4.7          \\ 
8  & 91  & 22.7   & 2.3   & 95.1           & & 36  & 4.1   & 2.4  & 39.5            & & 11  & 8.3  & 7.3  & \textit{25.4} \\ 
9  & 98  & 45.9   & 3.7   & \textit{879.7} & & 40  & 6.7   & 3.4  & \textit{399.7}  & & 6   & 11.6 & 10.2 & \textit{71.6} \\ 
10 & 99  & 206.1  & 7.8   & TL             & & 43  & 15.2  & 5.2  & \textit{1800.3} & & 12  & 15.4 & 13.7 & \textit{445.5}\\
11 & 99  & 697.9  & 20.6  & TL             & & 47  & 23.5  & 8.3  & TL              & & 9   & 21.5 & 18.9 & TL \\
12 & 100 & 2051.2 & 62.9  & TL             & & 52  & 39.2  & 13.3 & TL              & & 8   & 29.2 & 26.2 & TL \\
\hline
\end{tabular}
}
\label{find_all_table}
\end{table}

\section{Conclusions and Future Work}
  \label{conclusions}

We have proposed symmetry breaking predicates which can be added to the 
Boolean formula representing various DFA identification problems. 
By adding the predicates we can reduce the problem search space 
through enforcing DFA states to be enumerated in the depth-first 
search or the breadth-first search order. 

In the case of noiseless DFA identification we have compared 
translation-to-SAT method from~\cite{heule2010} to the same one
with proposed symmetry breaking predicates instead of original max-clique predicates.
The proposed approach clearly improved translation-to-SAT technique which 
was shown with the experiments on the randomly generated input data. 
The BFS-based approach have shown better result than the DFS-based 
one if the target DFA size is large enough.

We also have drawn our attention to the case of noisy DFA identification.
We have proposed a modification of the noiseless approach for the noisy case.
To achieve compact encoding for that case we have used the order encoding method.
We have shown that the previously proposed max-clique technique for 
symmetry breaking is not applicable in the noisy case while our BFS-based approach is. 
We have shown that the BFS-based strategy can be applied in the noisy case when 
an automaton which is consistent with a given set of strings does not exists. 
The current state-of-the-art EA from~\cite{lucas2005} cannot determine that. 
In experimental results, we have shown that our approach with BFS-based symmetry breaking predicates
clearly outperforms the algorithm without any predicates. 

We have proposed a solution for the find-all DFA problem. 
The proposed approach can solve the problem that the previously developed methods 
cannot be applied for efficiently.
We had performed the experiments which have shown that our approach
with the incremental SAT-solver clearly outperfoms the backtracking algorithm.

We plan to translate the problem of noisy DFA identification to Max-SAT in order to limit the number of corrections
without using an auxiliary array of integer variables. 
We also plan to experiment with alternative integer encoding methods.


\section*{Acknowledgements}

The authors would like to thank Igor Buzhinsky, Daniil Chivilikhin and Maxim Buzdalov for useful comments.
This work was financially supported by the Government of Russian Federation, Grant 074-U01,
and also partially supported by RFBR, research project No.~14-07-31337 mol\_a.

\section*{References}

\bibliography{main}

\end{document}